\documentclass[aps,prl,twocolumn,groupedaddress,showpacs,nofootinbib]{revtex4-1}
\newcommand{\beq}{\begin{equation}}
\newcommand{\eeq}{\end{equation}}
\newcommand{\beqa}{\begin{eqnarray}}
\newcommand{\eeqa}{\end{eqnarray}}
\usepackage{graphicx}
\usepackage{amsmath}
\usepackage{bm}
\DeclareMathOperator{\sech}{sech}

\begin{document}


\title{Shortcut to adiabatic passage in two and three level atoms}
%

\author{Xi Chen$^{1,2}$, I. Lizuain$^{1}$, A. Ruschhaupt$^{3}$, D. Gu\'ery-Odelin$^{4}$,
J. G. Muga$^{1}$}
\affiliation{
   $^{1}$ Departamento de Qu\'{\i}mica-F\'{\i}sica,
UPV-EHU, Apartado 644, 48080 Bilbao, Spain \\
   $^{2}$ Department of Physics, Shanghai University,
200444 Shanghai, P. R. China\\
   $^{3}$ Institut f\"ur Theoretische Physik, Leibniz
Universit\"{a}t Hannover, Appelstra$\beta$e 2, 30167 Hannover,
Germany\\
{$^{4}$Laboratoire Collisions Agr\'egats R\'eactivit\'e, CNRS UMR 5589, IRSAMC, Universit\'e Paul Sabatier, 118 Route de Narbonne, 31062 Toulouse CEDEX 4, France}
}


\date{\today}

\begin{abstract}
We propose a method to transfer the population and control the state
of two-level and three-level atoms speeding-up Adiabatic Passage techniques while keeping their robustness versus parameter variations. The method is based on
supplementing the standard laser beam setup of Adiabatic Passage methods with   auxiliary steering laser pulses of orthogonal polarization. This provides a shortcut to adiabaticity
driving the system along the adiabatic path defined by the standard setup.
\end{abstract}


\maketitle

{\it Introduction.---}Two major routes for manipulating the state of a quantum system with interacting  fields are based on resonant pulses
or on adiabatic methods, such as ``Rapid'' Adiabatic Passage (RAP), Stimulated Raman Adiabatic Passage (STIRAP) and their many variants. In general terms simple
fixed-area resonant pulses may be fast if intense enough, but quite unstable with respect to errors or fluctuations of the parameters, whereas adiabatic passage is robust but slow. For many applications, from Nuclear Magnetic Resonance (NMR)
to quantum information processing, the ideal method should be fast and robust, combining the best of the two worlds.
These two requirements are particularly demanding if
quantum computing is to become feasible at all.
It is possible to make the pulses more stable by combining them into pulse sequences, but in practice their use is limited by the longer times required with respect to the single pulse, the need to control phase angles and pulse durations accurately, or off-resonant excitations due to sharp pulse edges \cite{NMR}.
Moreover the error compensating properties of square-pulse sequences are not preserved when substituting them with smooth pulses so that
the design of good sequences requires ``a good portion of experience and magic'' \cite{Molmer}.
In NMR, composite pulses are increasingly superseded by adiabatic passage methods \cite{NMR}, which have also been very successful in chemical reaction dynamics \cite{Kral}, laser cooling \cite{lc}, atom optics \cite{RM}, metrology \cite{Salomon}, interferometry \cite{Chu}, or cavity quantum electrodynamics \cite{Lambro,Bergmann}. When robustness is the primary concern, they are quite sufficient, and have as well become basic operations for quantum information processing, either to design robust gates \cite{JJ,gates} or in quantum adiabatic computing \cite{AC1,AC2}, which relies on an adiabatic evolution of the ground state from an initial to a final Hamiltonian. If speed is also important, however, the limitations may be severe \cite{AC2}. Given the stated difficulties of composite pulses, it is then quite natural to look for  robustness and high operation velocities taking the adiabatic methods as the starting point and shortening their duration somehow. Our objective here is to propose a shortcut to adiabatic passage (abbreviated as ``SHAPE'' hereafter) using a recent formulation by Berry \cite{Berry09} of ``transitionless quantum driving'',
related to work by Kato \cite{Kato} on the adiabatic theorem.
The specific applications we shall discuss are
speeded-up versions of (2-level) RAP and (3-level) STIRAP
schemes, as canonical examples of other adiabatic methods. Variants such as fractional RAP or STIRAP, and multilevel schemes
may be treated along similar lines.

The philosophy of the transitionless quantum driving algorithm
\cite{Berry09} is to supplement the Hamiltonian $H_0(t)$
of a reference system
with an auxiliary term $H_1(t)$ to steer the dynamics exactly along the instantaneous eigenstates $|\lambda_n(t)\rangle$ of $H_0(t)$ without transitions among them,
formally in an arbitrarily short time,
%
\beq
H_1 (t)= i \hbar \sum_n(|\partial_t \lambda_n \rangle \langle \lambda_n |
 - \langle\lambda_n |\partial_t \lambda_n  \rangle | \lambda_n  \rangle\langle \lambda_n|).
%
\label{Berry's Hamiltonian}
\eeq
At variance with Lyapunov-control methods \cite{Wang},
the extra term is independent of the time-dependent state
so it leads to simpler, linear dynamics,
and moreover it provides systematically exact solutions for adiabatic following without the need for a trial and error
approach to find a good control field \cite{Wang}.
Regarding its physical realizability, in general
there is no guarantee that $H_1$ may be easy to implement and each case needs a separate study. For example, for an $H_0$ describing a particle in a time-dependent harmonic potential, $H_1$ turns out to be a non-local interaction, and its realizability in a useful parameter domain for cold atoms
remains an open question \cite{JPB2}.
(Transitionless dynamics for the harmonic oscillator may be achieved with a local interaction by
inverse engineering the frequency dependence with the aid of Ermakov-Riesenfeld invariants \cite{Xi}, or with state acceleration techniques \cite{Nakamura}.)
For a particle with spin in a time dependent magnetic field, $H_1$
becomes a complementary, time-dependent magnetic field \cite{Berry09}.
For the atomic two- and three-level systems studied here, $H_1$ will
involve laser interactions added to the original laser setup implied by $H_0$
as discussed below.

{\it Rapid Adiabatic Passage.---}Let us consider first the speeding-up of
a standard Rapid Adiabatic Passage that
inverts the population of two-levels of an atom,
$|0\rangle$ and $|1\rangle$, by sweeping the radiation through resonance.
This broadspread technique originated in Nuclear Magnetic Resonance \cite{Bloch}
but is used in virtually all fields where 2-level systems may be controlled by
external interactions, such as laser-chemistry, modern quantum optics or quantum information processing.
When the frequency sweep is much shorter than the life-time for
spontaneous emission and other relaxation times,
it is termed rapid adiabatic passage (RAP).

Using the rotating wave approximation, the Hamiltonian ${H}_0 (t)$ in a laser-adapted interaction picture can be written as
\beqa
\label{H0}
{H}_0 (t)= \frac{\hbar}{2} \left(\begin{array}{cc} \Delta (t)  & \Omega_{R}(t)
\\
\Omega_{R} (t)& - \Delta (t)
\end{array}\right),
\eeqa
where $\Omega_{R}(t)$ is the Rabi frequency, which we take to be real,
and $\Delta (t)= \omega_0- \omega_L$ the detuning,
assumed to change slowly on the scale of the optical period.
It is the difference between the Bohr transition frequency and the laser carrier frequency $\omega_L$,
due to a change in the carrier frequency or
a controlled alteration of the Bohr frequency by Zeeman or
Stark shifts.
%
The instantaneous eigenvectors are
\beqa
\label{instantaneuous states}
|\lambda_{+}(t)\rangle &=&  \cos[\theta(t)/2]
|1 \rangle - \sin[\theta(t)/2] |0 \rangle,\\
|\lambda_{-}(t)\rangle &=& \sin[\theta(t)/2]
|1 \rangle + \cos[\theta(t)/2]
|0 \rangle,
\eeqa
with the mixing angle $\theta (t)\equiv \arccos [-\Delta(t)/\Omega(t)]$
and  eigenvalues $E_{\pm}(t)= \pm \hbar \Omega /2$, where $\Omega = \sqrt{\Delta^2 (t) + \Omega^2_R (t)}$.
If the adiabaticity condition
\beq
\label{adiabatic condition}
\frac{1}{2}|\Omega_a| \ll |\Omega(t)|,
\eeq
where $\Omega_a\equiv[\Omega_R(t)\dot{\Delta}(t)-\dot{\Omega}_R(t)\Delta(t)]/\Omega^2$,
is satisfied,
the state evolving from $|\psi(t=0)\rangle=|\lambda_\pm(0)\rangle$ follows the adiabatic approximation
\beqa
\label{adiabatic states}
|\psi_{\pm}(t)\rangle= \exp{\left\{-\frac{i}{\hbar}\int^{t}_{0} dt' E_\pm(t')\right\}} |\lambda_\pm(t)\rangle,
\eeqa
whereas transitions will occur otherwise.
Different adiabatic passage schemes correspond to different specifications of
$\Omega_R$ and $\Delta$ for which $\psi_{\pm}$ passes from one bare
state to the other. The simplest one is the Landau-Zener scheme with
constant $\Omega_R$ and linear-in-time $\Delta$. For the examples below we shall use
the more adiabatic (and thus potentially faster) Allen-Eberly scheme \cite{AE,VG}:
$\Omega_R=\Omega_0\sech(\pi t/2 t_0)$, $\Delta=(2\beta^2 t_0/\pi) \tanh(\pi t/2 t_0)$.
Regardless of the scheme chosen ${H}_1(t)$ takes the form
\beqa
{H}_1 (t)
= \frac{\hbar}{2} \left(\begin{array}{cc} 0 & -i \Omega_{a}
\\
i \Omega_{a} & 0
\end{array}\right),
\eeqa
where (up to a phase factor)
$\Omega_a$ plays the role of the  Rabi frequency of the auxiliary field.
%
%
In principle $H=H_0+H_1$ drives the dynamics along the $H_0$-adiabatic path
in arbitrarily short times, but there are practical limitations
such as the laser power available.
Moreover, a comparison with $H_0$-dynamics is only fair if $|\Omega_a|$ is smaller or approximately equal to the peak Rabi frequency with the original laser setup.
Independently of the scheme chosen and in a range of interaction times that break down the adiabaticity condition,
it is remarkable that the dynamics can be driven along the $H_0$-adiabatic path
while fulfilling the inequalities $|\Omega_{a}|\le|\Omega|\le |\Omega_0|$.

The physical meaning and realizability of the auxiliary term
are determined by going back to the Schr\"odinger picture: it represents
a laser with the same time dependent frequency of the
original one, but a differently shaped time-dependent intensity,
and perpendicular polarization.

\begin{figure}[t]
\begin{center}
\includegraphics[width=0.60\linewidth]{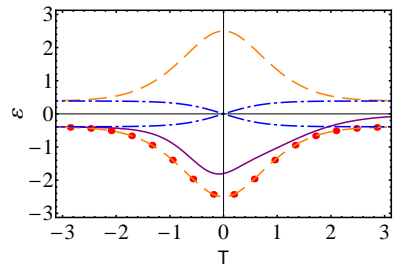}
\end{center}
\caption{\label{figAE} (color online). (Dimensionless) energies $\varepsilon=E/(\beta \hbar)$
versus (dimensionless) time $T=t\beta$ for the AE scheme: Diabatic energies, dot-dashed (blue) lines; adiabatic energies, dashed (orange) lines; average energy evolving with $H_0$, solid (purple) line; average energy evolving with
$H=H_0+H_1$, dotted (red) line, indistinguishable from the lower adiabatic energy.
$\omega=5$, $\tau=1.22$.}
\end{figure}

%
%
For the Allen-Eberly scheme the population of the excited state $P_1$
starting from the ground state depends
on the dimensionless parameters
$\tau=t_0\beta$ and $\omega=\Omega_0/\beta$ \cite{VG}:
$P_1= 1 - \sech^2(2 \tau^2/\pi)\cos^2[\tau  (\omega^2 - 4 \tau^2/\pi^2)^{1/2}]$.
A population transfer near to one ($P_1>0.999$) and stable versus parameter variations is achieved for $\omega\ge 3$ and $\tau\ge 3$. We may calculate $\Omega_a$ and the minimal time for which the maximum of $\Omega_a$ with respect to $t$ is $\le\Omega_0$. In the stated range
this is accurately given by $\tau_m=\pi/(4\omega)$, or
%
$t_{0,m}=\pi/(4 \Omega_0)$.
%
The reduction factor with respect to the adiabatic time $\tau_a\approx 3$  may be very  significant, $t_{0,m}/t_a\approx{\pi}/{(12\omega)}$, this is
$0.09$ for $\omega=3$, or $0.01$ for $\omega=20$.
Of course the SHAPE Hamiltonian $H$ may also drive the system along the adiabatic
path outside the $w,\tau>3$ domain as illustrated in Fig. \ref{figAE}.
For comparison, the population of the excited state due to a square $\pi$ pulse
with on-resonance Rabi frequency $\Omega_0$ is
\beqa
P_{1} = \frac{\Omega_0}{\Omega}\left|\sin \left(\frac{\Omega t}{2}\right)\right|^2.
\eeqa
Complete population transfer requires $\Delta=0$, and
a pulse time
%
$t_R= \frac{\pi}{\Omega_{0}}$.
%
For the same $\Omega_0$ and limiting the auxiliary
laser by $|\Omega_a|\le \Omega_0$, the minimal characteristic
time $t_{0,m}$ of the SHAPE method is of the order of $t_R$, $t_{0,m}=t_R/4$.
In fact the actual interaction time to implement a successful population inversion  with the AE scheme (SHAPE corrected or not) should be a few times $t_{0}$; this may be estimated from the
dependence of the excited population of the adiabatic state with time, which is
$>.999$ for $t\ge 8 t_{0}$.

Figures 2 and 3 show examples of the fidelity ($P_1$) with respect to variations in the Rabi frequency and detuning  with SHAPE (AE scheme for $H_0$), the evolution with $H_0$ (AE scheme), a Rabi $\pi$-pulse, and a composite $\frac{\pi}{2}(x)\pi(y)\frac{\pi}{2}(x)$ pulse, a fault-tolerant combination where $x,y$ refer to the laser polarization (and Pauli matrix) involved.
Clearly SHAPE provides a fast, robust and efficient population inversion compared to all other methods. All cases are for the same $\Omega_0$, and in SHAPE  $|\Omega_a|\le \Omega_0$.
%
%
\begin{figure}[ht]
\begin{center}
\includegraphics[width=0.48\linewidth]{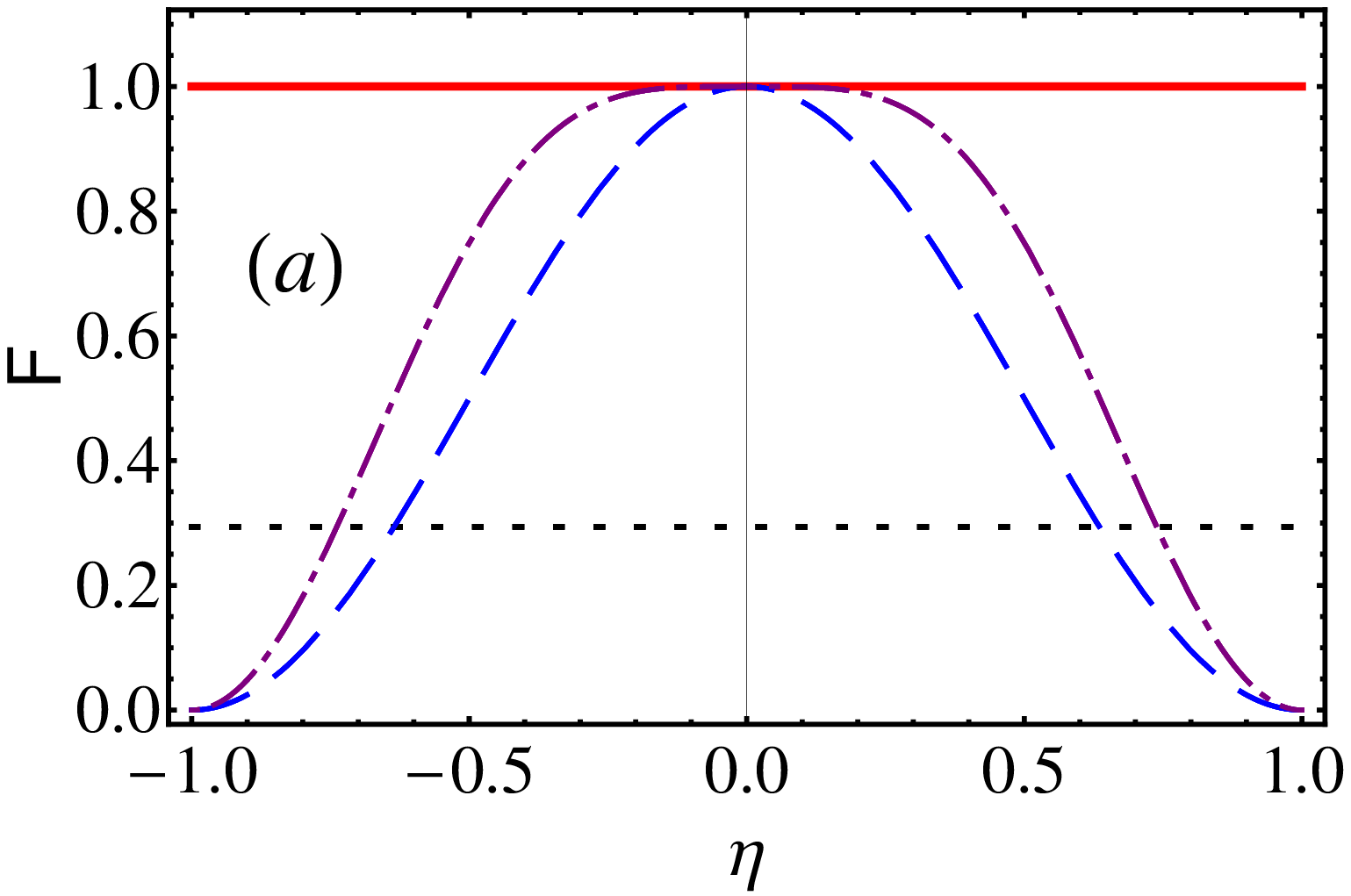}
\includegraphics[width=0.48\linewidth]{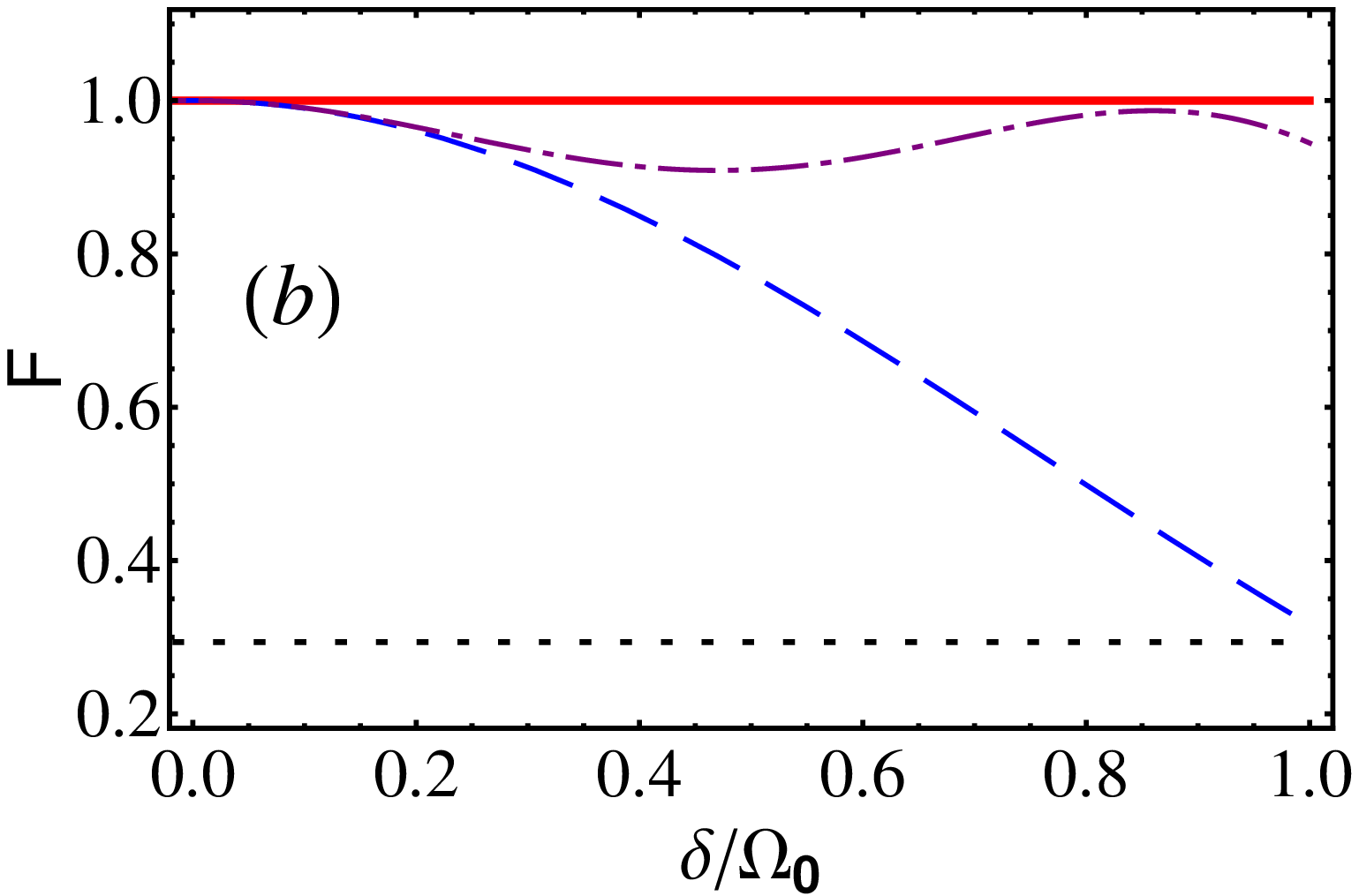}
\end{center}
\caption{\label{fig.3} (color online). Dependence of the fidelity ($P_1$) on the changes of (a) the Rabi frequency from $\Omega_0$ to $\Omega_0(1+\eta)$ (b): the detuning by
$\delta$.
SHAPE, AE scheme (solid, red line) with $\Omega_a\le\Omega_0$;
ordinary Adiabatic Passage
(dotted, black line); Rabi pulse (dashed, blue line);
Composite pulse $\frac{\pi}{2}(x)\pi (y)\frac{\pi}{2}(x)$ (dot-dashed, purple line).
$\Omega_0=2 \pi\times 5$ MHz; $t_0=25$ ns; $\beta=2\pi$ MHz.}
\end{figure}
%

{\it Stimulated Raman Adiabatic Passage.---}Similar ideas can be applied to three-level STIRAP. The
Hamiltonian ${H}_0 (t)$ for the two-photons resonance case
within the rotating wave approximation (RWA) and in a laser adapted interaction picture reads \cite{Bergmann}
\beqa
{H}_0 (t)= \frac{\hbar}{2} \left(\begin{array}{ccc} 0 & \Omega_{p}(t) & 0 \\ \Omega_{p}(t) & 2 \Delta & \Omega_{s}(t)
\\
0 & \Omega_{s}(t) & 0
\end{array}\right),
\eeqa
in terms of the Rabi frequencies for the Stokes, $\Omega_{s}(t)$,
and pumping lasers, $\Omega_{p}(t)$,
and the laser detuning $\hbar \Delta= (E_2-E_1)- \hbar \omega_p=(E_2-E_3)- \hbar \omega_s$.
The instantaneous eigenstates $|\lambda_n \rangle$ are
\beqa
\label{instantaneuous states-for 3-level system}
\nonumber
|\lambda_{+} (t) \rangle &=&  \sin \theta \sin \phi |1\rangle + \cos \phi |2\rangle + \cos \theta \sin \phi |3\rangle, \\ \nonumber
|\lambda_{-} (t) \rangle &=&   \sin \theta \cos \phi |1\rangle  - \sin \phi |2\rangle + \cos \theta \cos \phi |3\rangle,
\\
|\lambda_{0} (t) \rangle &=&  \cos \theta |1\rangle  - \sin \theta |3\rangle,
\eeqa
with eigenvalues given by $E_{+} (t)=\hbar \Omega \cot(\phi/2)$, $E_{0}=0$, and $E_{-} (t) = -\hbar \Omega \tan(\phi/2)$.
The time-dependent mixing angles $\theta$ and $\phi$ are respectively defined by $\tan \theta = \Omega_{p} (t)/ \Omega_{s} (t)$ and $\tan (2\phi) = \Omega / \Delta(t)$, whereas
$\Omega =\sqrt{\Omega^2_{p}(t) + \Omega^2_{s}(t)}$.
The population transfer $1\rightarrow 3$
is realized by the ``dark state'' $|\lambda_0\rangle$.

\begin{figure}[ht]
\begin{center}
\scalebox{0.34}[0.34]{\includegraphics{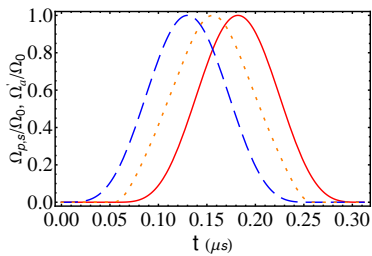}}
\end{center}
\caption{\label{fig.4} (color online). Time evolution of Rabi frequencies in a STIRAP
sequence of laser pulses defined by Eq. (\ref{oms}) with
$\Omega_0 = 2 \pi \times 5$ MHz, $\Delta = 2 \pi  \times 0.5$ MHz, and ${\sf T} = 0.26$ $\mu$s.
Dashed blue line: $\Omega_s/\Omega_0$;  Solid red line: $\Omega_p/\Omega_0$; The dotted orange curve represents $\Omega'_a/\Omega_0$.}
\end{figure}

\begin{figure}[ht]
\begin{center}
\scalebox{0.28}[0.28]{\includegraphics{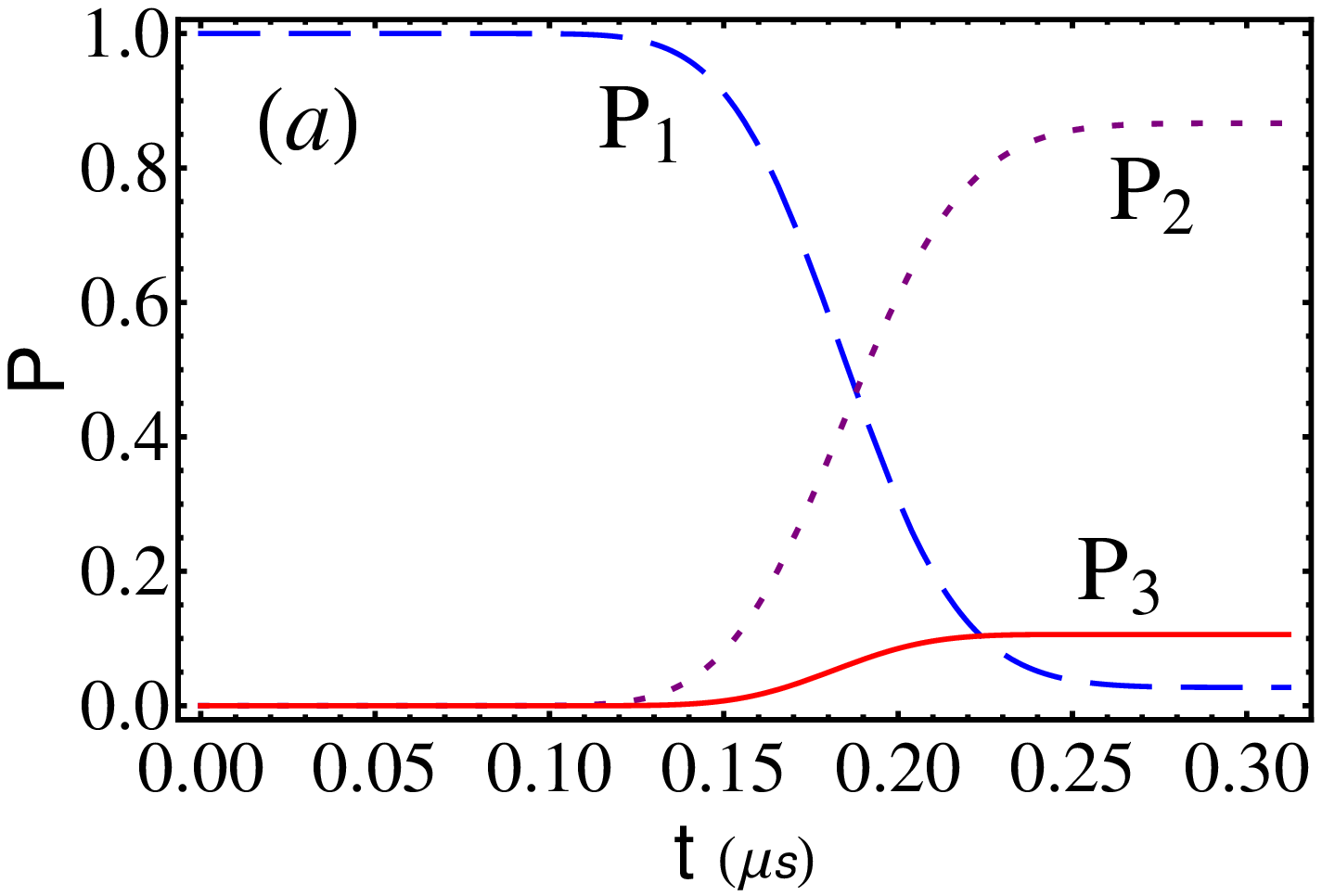}}
\scalebox{0.28}[0.28]{\includegraphics{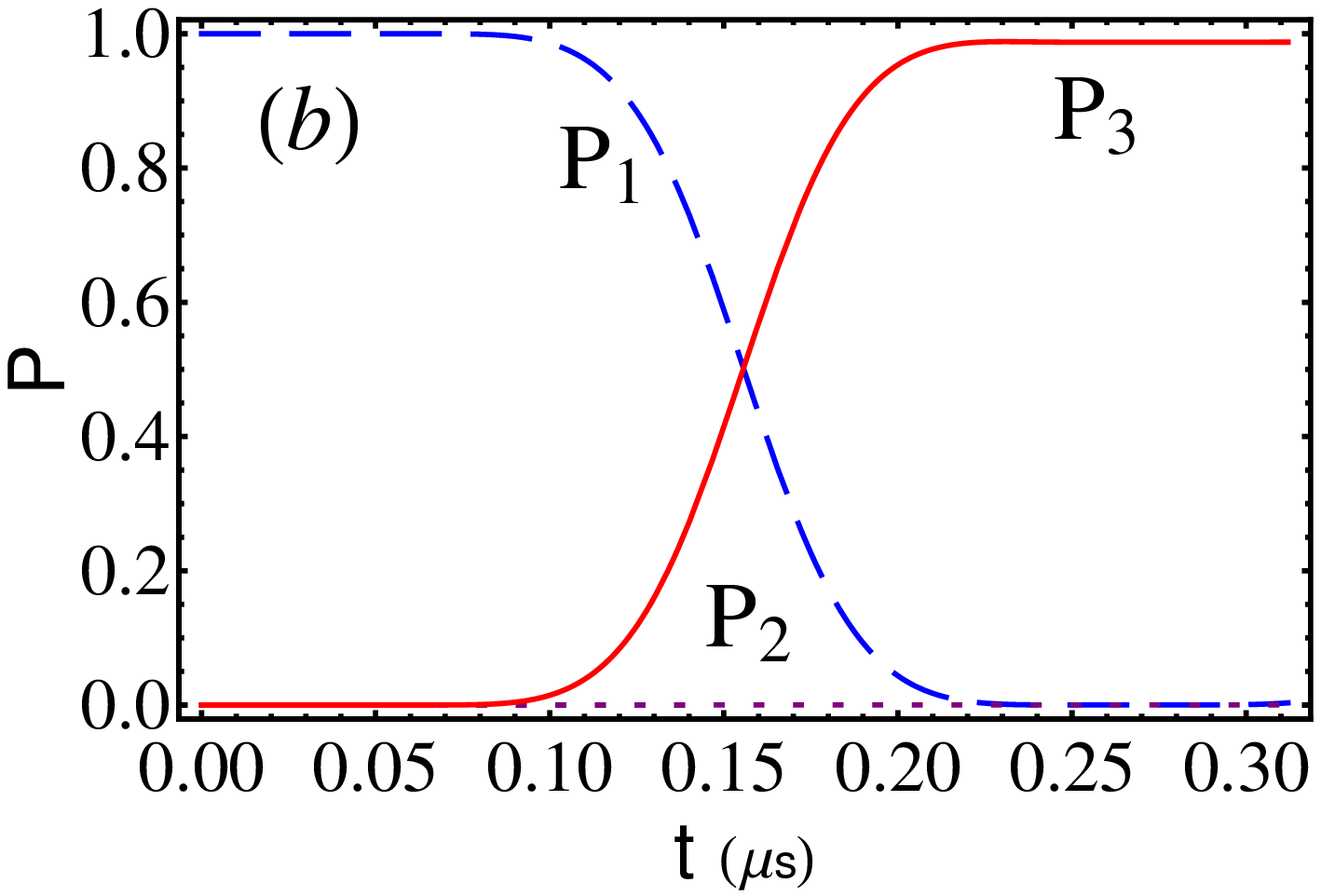}}
\end{center}
\caption{\label{fig.5} (color online). Time evolution of the populations of levels 1, 2 and 3 for STIRAP: (a) Hamiltonian ${H}_0 (t)$;
(b) Hamiltonian ${H} (t)$ with additional ${H}_1 (t)$. Same parameters as in Fig. \ref{fig.4}.}
\end{figure}

The Hamiltonian ${H}_1 (t)$,
takes now the form
\beqa
\label{Berry3}
{H}_1 (t)= i \hbar \left(\begin{array}{ccc} 0 &\dot{\phi}\sin \theta  &  \dot{\theta}
\\
-\dot{\phi} \sin \theta  & 0 &  -\dot{\phi}\cos \theta
\\
-\dot{\theta} & \dot{\phi}  \cos \theta& 0
\end{array}\right),
\eeqa
with
$
\dot{\theta}= [{\dot{\Omega}_{p} (t) \Omega_{s} (t)- \dot{\Omega}_{s} (t) \Omega_{p} (t)}]/{\Omega^2},
$
and
$
\dot{\phi}=[{(\dot{\Omega}_{p} (t) \Omega_{p} (t) + \dot{\Omega}_{s} (t) \Omega_{s} (t)) \Delta (t) }]/[{2 \Omega (\Delta^2+\Omega^2)}].
$
We would need in principle three new lasers to implement this Hamiltonian.
The ones connecting levels $1$-$2$ and $2$-$3$ should have the same frequency
as the original ones but orthogonal polarization,
and the field connecting levels $1$-$3$ should be on resonance with this transition
to get an interaction picture Hamiltonian like (\ref{Berry3}) (The RWA approximation
is assumed in all cases.)
If $1$-$3$ is electric-dipole-forbidden,
a magnetic dipole transition may be used instead.
If we are only interested in performing a full passage from 1 to 3 and do not
want to reproduce all the effects of the full Hamiltonian $H_0+H_1$, $H_1$ may be simplified by retaining just the $1$-$3$ interaction,
\beq
{H}'_1 (t)= \frac{\hbar}{2} \left(\begin{array}{ccc} 0 & 0  & i\Omega'_a
\\ 0 & 0 &  0  \\
-i \Omega'_a & 0 & 0
\end{array}\right),
\label{13}
\eeq
where $\Omega'_a = 2 \dot{\theta}$.
That this is so may be seen by working out the Schr\"odinger equation in the adiabatic
basis: $d\langle \lambda_0(t)|\psi(t)\rangle/dt$ does not depend on $\dot{\phi}$ so that the $1$-$3$ and $2$-$3$ auxiliary lasers can be left out without affecting
$\langle\lambda_0(t)|\psi(t)\rangle$.
In the examples below we have chosen the pulse shapes \cite{Fewell}
\beqa
\nonumber
\Omega_{p}(t) &=& \Omega_{0}(t) f(t-\tau) ;~~ \Omega_{s}(t) =  \Omega_{0}(t) f(t),
\\
\label{oms}
f(t) &=& \left\{
\begin{array}{ll}
\sin^4(\pi t/{\sf{T}}) ~~ &(0 < t < {\sf T}), \\
0 ~~~~~&(\mbox{otherwise).}
\end{array}
\right.
\eeqa
Fig. \ref{fig.4} shows a STIRAP Stokes-Pump pulse sequence
where ${\sf T}$ is too short for complete population transfer because adiabaticity breaks down,
see Fig. \ref{fig.5}a.
We can remedy that with the auxiliary interaction in (\ref{13}), see
Figs. \ref{fig.4} and \ref{fig.5}b.
Keeping $|\Omega_a'|\le \Omega_0$ the process duration
is reduced approximately ten times with respect to the ordinary STIRAP scheme.
%

{\it Discussion and conclusions.---}A method to achieve fast and robust population transfer in two-level and three-level atomic systems has been presented,
based on supplementing the laser setup of standard adiabatic passage methods (RAP or STIRAP) by additional, properly time-shaped pulses with orthogonal polarization.
The Hamiltonian $H_1(t)$ that describes the additional steering pulses providing a shortcut to adiabaticity is given by a general algorithm to drive quantum systems without transitions \cite{Berry09}.
Other states (such as superpositions of two-levels) may be prepared by
speeded-up (SHAPE) versions of fractional RAP or STIRAP, and, if necessary, the phases may be controlled thanks to the freedom to choose the reference Hamiltonian $H_0(t)$ or phase gates.
As an outlook, similar techniques may provide a way to carry out adiabatic computation in a finite time \cite{Aha,bra}, or to speed up logic gates
based on adiabatic processes \cite{CBZ,Ion}, interferometric
techniques in superconducting qubits \cite{Shevchenko} or quantum dots \cite{Jong}, and the creation of entangled pairs of two-state systems \cite{Unanyan}.
The SHAPE method is compatible with approaches that optimize $H_0$ such as the quantum brachistochrone approach \cite{Zan}, since, after optimizing the adiabatic process, it leads to the design of even faster process.
Other adiabatic techniques \cite{SCRAP} may also benefit from these ideas
speeding up the avoided level crossings and keeping their stability versus parameter
variations.

We thank M. V. Berry and J. H. Eberly for discussions, and
acknowledge funding by Projects No. GIU07/40, No. FIS2009-12773-C02-01,
No. NSFC 60806041, No. 08QA14030, No. 2007CG52, No. S30105,
No. ANR-09-BLAN-0134-01, and Juan de la Cierva Program.

\end{document}